\documentstyle[preprint,aps]{revtex}

\tighten
\begin{document}
\draft

\title{\flushleft On Bohr's response to the clock-in-the-box 
thought experiment of Einstein\footnote{This work was conducted 
independently of official duties in the Centers
for Disease Control and Prevention (CDC), and the views do not 
necessarily represent those of the CDC.}} 
\author{\flushleft V Hnizdo}

\address{\flushleft
National Institute for Occupational Safety and Health,
1095 Willowdale Road, Morgantown, WV 26505, USA}

\address{\flushleft\rm
{\bf Abstract}. The recent analysis of De la Torre, Daleo and Garc\'{\i}a-Mata
of the reply of Bohr to the famous clock-in-the-box challenge of 
Einstein is criticized.}

\maketitle

\section*{}
\noindent 
Recently, De la Torre, Daleo and Garc\'{\i}a-Mata (TDG) \cite{TDG} 
have subjected to a `careful and irreverent' analysis the arguments 
of Einstein and Bohr in their dialog concerning the famous 
clock-in-the-box {\it gedanken} experiment devised by Einstein in an 
attempt to disprove the uncertainty relation for energy and time \cite{B},
and reached the conclusion that `Einstein's argument is flawed while 
Bohr's reply is wrong.' In the present note, we would like to point out that
it is the TDG analysis of the balancing procedure assumed in Bohr's reply 
that is flawed, and to uphold Bohr's simple inequality characterizing 
this balancing as correct---if not `obviously', then on a little reflection.

The inequality in question,
\begin{equation}
\Delta  p <T g \Delta m
\end{equation}
puts an  upper bound on the uncertainty $\Delta p $ of the momentum
of a box whose mass 
has been determined to an uncertainty $\Delta m$ using a spring balance
in a balancing procedure taking  a time $T$; 
$g$ is here the acceleration due to gravity. The rest of this paragraph 
is an elaboration on Bohr's terse justification of this 
inequality. Weighing on a spring balance, as on any balance, 
is essentially a position measurement (but also the preparation 
of a suitable  momentum state, a point to which we shall return 
below) in which the position of a pointer attached to the
mass being weighed is determined against a scale fixed firmly
to the reference frame. The balancing procedure of TDG, in agreement with
that implied by Bohr, consists of adding to or removing from
the box progressively smaller masses $\Delta m$ until the pointer
coincides with a reference mark on the scale to within some latitude
$\Delta q$.
Now, according to the uncertainty principle,
the uncertainty $\Delta p$ in the momentum state of the box then will be
of the order of $\hbar/\Delta q$, but this uncertainty
has an  upper bound due to the  fact that the driving force of the 
final balancing stage is the weight $\Delta m\,g$ of the last 
mass increment 
(or decrement) $\Delta m$, and thus it cannot impart to the box an impulse
greater than $\Delta m\,gT_{\Delta m}$, where $T_{\Delta m}$ is the time 
that the final stage takes to complete.
If this is not seen as `obvious', a simple classical energy-balance calculation
bears it out immediately. Consider a mass $m$ suspended in the state of
rest  from 
a spring of spring constant $k$, and suppose that a mass $\Delta m$ is added
to it. Assuming for the moment no damping, the sum of the elastic, 
gravitational and kinetic energies at
a distance $q$, measured vertically down from the original equilibrium
position, remains equal to the initial zero total energy:
\begin{equation}
\case{1}{2}k q^2-\Delta m\,gq+\case{1}{2}(m+\Delta m)v^2= 0.
\end{equation}
Here, only the gravitational energy of the increment $\Delta m$ need
be considered, as the weight of the original mass $m$ is balanced 
by the spring force at the original equilibrium position $q=0$.
The substitution in (2) of the value $q=\Delta m\,g/k$, which is where a 
spring force $-kq$ balances 
the weight $\Delta m\,g$ of the mass increment and the system reaches the 
maximum  speed $v=v_{\rm max}$, yields for the maximum momentum $p_{\rm max}$
the value
\begin{equation}
p_{\rm max}=(m+\Delta m)v_{\rm max}=\Delta m\,g
\sqrt{\frac{m+\Delta m}{k}} =\Delta m\,g\frac{\tau}{2\pi} 
\end{equation}
where $\tau$ is the oscillation period of the system.
Now, it takes at least a time $T_{\Delta m} \sim\tau$ 
to bring the system to the state of approximate rest at the new equilibrium 
position by reducing suitably its oscillations,
and so we have indeed
\begin{equation}
\Delta p\lesssim p_{\rm max}=\Delta m\,g\frac{\tau}{2\pi}
<\Delta m\,gT_{\Delta m}.
\end{equation}
As $T_{\Delta m} < T$, Bohr's inequality (1),  where $T$ is the  
time available for 
the whole balancing procedure, will be thus satisfied to an ample margin.

Contrary to what TDG assert, damping of oscillations
leading to dissipation of the kinetic energy of the box can be allowed.
In the weighing of the box before the release of a `photon',
a transfer of energy to the box  is not critical.
In the weighing after the release, however, it is permissible 
to transfer only an energy much smaller than $\Delta m\,c^2$ 
to the internal energy of the box.  
A complete damping of the oscillations  dissipates 
an energy $K_{\Delta m}=\case{1}{2}p_{\rm max}^2/(m+\Delta m)$, which,
according to equation (3), equals $(\Delta m\,g)^2/2k$, and so
the condition $K_{\Delta m}\ll \Delta m\,c^2$ can be written 
in terms of the new equilibrium position $q=\Delta m\,g/k$
as $q\ll 2c^2/g\sim 10^{16}$ m---and surely that  would be satisfied
more than amply in any meaningful weighing procedure, {\it gedanken} 
or `real'. In any case,  the transfer of heat 
energy to the box can be prevented by using as the damping mechanism 
the friction between the spring, suitably shaped for this purpose, 
and a medium, and conducting the generated heat energy from the spring 
to the environment of the whole system.  
Oscillations can be reduced also by adding the given mass increment
$\Delta m$ to the box gradually---for example, by breaking $\Delta m$
into suitably small parts and adding these to the box one by one
in a suitable time sequence (classically,  oscillations are 
eliminated when a mass increment $\Delta m\ll m$ is brought onto the box 
linearly in a time equal to the period $\tau=2\pi(m/k)^{1/2}$, see \cite{LL}). 

The above considerations stem from the realization that the completion of 
a proper weighing procedure
amounts to the preparation of a state of approximate rest in which  
momentum as well as position must be suitably determined. This important point
seems to be overlooked in most criticisms of Bohr's response, but it has been
emphasized by Peierls, who treats the weighing procedure
essentially as a momentum measurement \cite{Pei}.

The upper bound (1) on the momentum uncertainty, together with the 
uncertainty relation $\Delta p\Delta q\sim \hbar$, means that 
the uncertainty $\Delta q$ in the box's position has a lower bound,
\begin{equation}
\Delta q > \frac{\hbar}{Tg\Delta m}.  
\end{equation}
Bohr comments this by saying \cite{B}: `The greater the accuracy  
of the reading $q$ of the pointer, the longer must, consequently, 
be the balancing interval $T$, if a given accuracy $\Delta m$ of the
weighing of the box with its content shall be obtained.'
This should not be misunderstood as meaning that the increase of
the balancing time $T$ brought about by just a decrease in the damping
of the oscillations would permit a smaller uncertainty $\Delta q$,
since it is the free-oscillation period $\tau$,  
and that is controlled by the spring constant $k$, that
determines the maximum possible
momentum  uncertainty $\Delta p\sim \hbar/\Delta q$ for a given
value of $\Delta m$ (see equation (4)).
Bohr used inequality (5) and 
general relativity's `red-shift' formula $\Delta T=Tg\Delta q/c^2$
to confirm the energy--time uncertainty relation 
$\Delta E\Delta T>\hbar$
for the uncertainty $\Delta E=\Delta m\,c^2$ in the energy of a 
`photon' and the uncertainty $\Delta T$ in the timing of its release from the
box by a momentary opening of a shutter operated by a clock in the box.
TDG claim that $\Delta T$ is in Bohr's usage
the indeterminacy in the balancing time $T$ of the box, despite
Bohr's statement \cite{B} that the clock `will change its rate in
such a way that its reading in the course of a time interval $T$ will 
differ by an amount $\Delta T$.' This clearly means that the timing 
of the opening of the shutter operated by the clock will be 
uncertain to within the amount $\Delta T$ because the latter is,  
according to the `red-shift' formula, directly proportional
to the displacement of the clock from the reference mark
with respect to which the clock was synchronized with the laboratory 
time,\footnote{
The distinction between the internal time of the clock in the box and 
the external laboratory time is important here, see \cite{AR}.} and
that displacement is uncontrollable within the latitude $\Delta q$.
Thus Bohr's  meaning of $\Delta T$
is the same as that assigned to it by Einstein, contrary
to what TDG assert. 

TDG attempt to refute (1) by a counterexample in which the box
is described by a harmonic-oscillator coherent state; in such states
the momentum uncertainty is fixed at 
$\Delta p=(\frac{1}{2}\hbar m\omega)^{1/2}$
and the energy uncertainty is given by $\Delta E=\hbar \omega |\alpha|$,
where $\omega$ and $\alpha$ are the oscillator frequency and  an 
arbitrary complex parameter, respectively.
Putting $\Delta E=\Delta m\,c^2$, and choosing $|\alpha|<
(c^2/Tg)(m/2\hbar\omega)^{1/2}$, where $T$ is now an arbitrary time parameter,
TDG obtain immediately
a momentum uncertainty $\Delta p> Tg\Delta m$, which would contradict (1).
However, such reasoning is flawed on at least two accounts. First, it
identifies the uncertainty in the mass of the box itself with that of
the oscillator's total bound-state energy $E$, 
which is the sum of the kinetic and elastic potential energies of the box. 
Second, it ignores the actual time needed to
bring the box into the desired state in which its position remains
within the given latitude $\Delta q\sim\hbar/\Delta p$  on the 
reference mark of the fixed laboratory scale. Inequality (1) simply says 
that a time $T>\Delta p/g\Delta m$ is needed 
for the preparation of such a state in a balancing procedure that involves 
a force $\Delta m\,g$, and nothing in the argument of TDG disproves 
the validity of this statement.
Thus if the TDG identification $\Delta m=\hbar\omega|\alpha|/c^2$
were correct, the balancing procedure resulting in   
a coherent harmonic-oscillator state with the desired properties would require
a time $T>\Delta p/g\Delta m=(c^2/|\alpha|g)(m/2\hbar\omega)^{1/2}$---and that
directly contradicts the TDG choice of the value of the 
parameter $|\alpha|$. 

The time involved in the preparation of a state is
of importance here, not the magnitude of the state's momentum 
spread relative to the spread in total energy.
One can find easily quantum states in which the momentum spread $\Delta p$
is greater by any given suitably dimensioned factor than
the spread $\Delta E$ in the total energy of the state, 
a trivial example being any stationary bound state, in which, 
by definition, $\Delta E=0$, but $\Delta p\ne 0$. 
For the TDG argument, the most favourable value of the parameter $\alpha$ 
is $\alpha=0$, which gives the stationary ground state of a harmonic 
oscillator.  Suppose, for the sake of the argument, that the box is 
prepared in such a state and that the readings $q$ on 
the balance scale are calibrated in terms of the mass $m$
as $kq=mg$. Then, despite the vanishing total-energy-uncertainty $\Delta E$
in the given state, the uncertainty $\Delta m$ to which 
the box's mass is determined is nonzero, satisfying the relation
\begin{equation}
\Delta m\,g\approx k\Delta q=\frac{k\Delta p}{m\omega}
=\frac{2\pi}{\tau}\Delta p
\end{equation}
where we used the relation $\Delta q=\Delta p/m\omega$ 
between the position and momentum uncertainties in a stationary
state of a harmonic oscillator of
frequency $\omega=2\pi/\tau=(k/m)^{1/2}$. 
Relation (6) was obtained  
without any consideration given to the time involved in the preparation of 
the quantum state concerned---but when the oscillator period 
$\tau$ is accepted as a lower bound on the time needed to prepare the state,
relation (6) agrees with relation (4), which was obtained
classically for a balancing procedure explicitly involving
time. However, it should be noted here that
the fact that the box is a {\it macroscopic} system
means that its preparation in any {\it pure} 
quantum state, and the TDG counterexample is such a state,
would face fundamental difficulties even if the preparation
procedure was not restricted to the balancing procedure
considered.

Before closing, we would like to make two additional remarks.
The first one is on the emphasis of TDG on the need to distinguish between 
classical `uncertainty',
by which they mean `the lack of precision in the knowledge of a value
assigned to an observable due to apparatus or experimental limitations',
and quantum `indeterminacy', which `denotes the impossibility of 
assigning precise values to the observables of a system as prescribed 
by quantum mechanics;' TDG reproach Bohr for mixing confusingly 
the two notions in his argumentation. 
In this connection, TDG point out that 
quantum `indeterminacies' of
observables are not related by the same function that relates the
observables themselves. But quantum `indeterminacies' are not different
in this respect from experimental `uncertainties'. To use the example
of the energy $E$ and momentum $p$ of a free
particle, where $E=p^2/2m$: when the mean momentum $|\bar{p}|\gg\Delta p$, 
the `indeterminacy'  
$\Delta E\approx |dE/dp|_{p=\bar{p}}\Delta p=(|\bar{p}|/m)\Delta p$,  
exactly as for an experimental `uncertainty' 
$\delta E\approx(|p|/m)\delta p$. While it must be admitted that
it is often necessary to make a distinction between 
`uncertainty' and `indeterminacy',  these two concepts
are closely related in the context of establishing 
the empirical meaning of the quantum-mechanical formalism, which
is the context in which Bohr uses such terms. According to Bohr,
quantum mechanics would not be a satisfactory physical theory
if it were not possible to demonstrate that there is 
harmony between what he called the possibilities  of definition
in the formalism on the one hand and the physical possibilities of 
measurement on the other \cite{AP} (the most elaborate
embodiment of this point of view of Bohr, as applied to the case
of quantum electrodynamics,  is the famous 
analysis of the measurability of the electromagnetic field of
Bohr and Rosenfeld \cite{BR}). There is such harmony only
if one cannot prepare experimentally 
a state in which a physical quantity $Q$ has an `uncertainty' 
$\delta Q$ while, according to quantum mechanics, that quantity
can be defined in such a state only to an `indeterminacy' 
$\Delta Q> \delta Q$, and, conversely, if the formalism does not
allow a sharper definition of the values of physical quantities 
than that permitted by the measurement (albeit perhaps only
{\it gedanken}) possibilities.  
Bohr believed that quantum mechanics is a satisfactory
physical theory, and therefore that the `uncertainty'
$\delta Q$ of an experimental preparation must match  
the `indeterminacy' $\Delta Q$ 
in all the physical quantities of a physical state; 
it is in this light that Bohr defended the uncertainty relations against
the ingenious attempts of Einstein to disprove them.
After Bohr's criticism of the clock-in-the-box thought experiment, Einstein
abandoned his attempts at showing that simultaneous {\it determination} of two 
conjugate quantities is in principle experimentally possible, and instead began 
to argue `only' for simultaneous {\it reality} of such quantities.
The famous argument of Einstein, Podolsky and Rosen \cite{EPR} that
the quantum-mechanical formalism cannot provide  a complete description
of physical reality is constructed in this spirit.   

The second remark concerns the point of TDG that Einstein's
use of the photon as the particle with which he aimed to disprove 
the energy--time uncertainty relation was unfortunate,
rendering his argument flawed.
Indeed, already according to classical electrodynamics,
an opening of the shutter for a short time $\Delta t$
would result in the emission of an
electromagnetic-wave pulse containing frequencies in a 
range of width $\Delta \omega \sim 1/\Delta t$, but this is a classical
Fourier-analysis result that alone does not lead to
quantum uncertainty relations (a classical 
electromagnetic pulse containing many frequencies still has
a well-defined energy). Now, the photon, {\it as a 
quantum particle}, cannot be sharply localized in 
time and simultaneously have a sharply defined energy,
just as a molecule, {\it as a quantum particle}, 
cannot be sharply localized in time 
and have a sharply defined kinetic energy. Nothing of substance
would change in the clock-in-the-box experiment if
Einstein used in it a gas of material particles like molecules 
instead of the electromagnetic radiation.\footnote{This is why
we prefer, following Peierls \cite{Pei} and Pais \cite{Pais},
to employ the term `clock' rather than `photon' in the name 
we use for this {\it gedanken} experiment.} 
Essentially, the clock-in-the-box {\it gedanken} experiment  
boils down to the question whether it is possible, 
in a well-defined (though not necessarily realistic) measurement 
procedure, to determine arbitrarily 
accurately both a change in the energy of a system
and the laboratory time when the change takes place,
irrespective of the particular form the energy change
assumes---and that may be the emission of electromagnetic radiation, 
matter, or both. To argue, as TDG do, that Einstein assumed
a non-existing particle, namely a well-localized photon of
a well-defined energy, amounts to `putting the cart before 
the horse', because while such a particle  cannot be defined 
in the formalism of quantum theory, it is up to  experiment 
(here, a {\it gedanken} one) to check
whether that formalism is not too narrow for the scope of physical
possibilities of measurement.

\end{document}